\documentclass[11pt]{article}
\usepackage{fullpage,times,Stdinclude} 
\usepackage{epsfig}
\usepackage{graphicx}

\begin{document}

\title{Towards a Theory of Cache-Efficient Algorithms
       \thanks{
       Some of the results in this appeared in a preliminary form in
       the Proceedings of the Eleventh ACM-SIAM Symposium on Discrete
       Algorithms 2000 \cite{SC1}.\newline
       This work is supported in part by DARPA Grant DABT63-98-1-0001,
       NSF Grants CDA-97-2637 and CDA-95-12356, The University of
       North Carolina at Chapel Hill, Duke University, and an
       equipment donation through Intel Corporation's Technology for
       Education 2000 Program.  The views and conclusions contained
       herein are those of the authors and should not be interpreted
       as representing the official policies or endorsements, either
       expressed or implied, of DARPA or the U.S. Government.}  }

\author{
Sandeep Sen\thanks{
Department of Computer Science and Engineering, IIT Delhi, New
Delhi 110016, India.
E-mail:\texttt{ssen@cse.iitd.ernet.in}.
Part of the work was done when the author was a visiting faculty in 
the Department of Computer Science, University
of North Carolina, Chapel Hill, NC 27599-3175, USA.
}
\and
Siddhartha Chatterjee\thanks{
        Department of Computer Science,
        University of North Carolina, Chapel Hill, NC 27599-3175, USA.
        E-mail: \texttt{sc@cs.unc.edu}.}
\and
 Neeraj Dumir\thanks{
Department of Computer Science and Engineering, IIT Delhi, New
Delhi 110016, India.}}

\date{}

\maketitle
\thispagestyle{empty}


\begin{abstract}
We describe a model that enables us to analyze the running time of an
algorithm in a computer with a memory hierarchy with limited
associativity,
in terms of various cache parameters.
Our model,
an extension of Aggarwal and Vitter's I/O model,
enables us to establish useful relationships between the cache
complexity and the I/O complexity of computations.
As a corollary,
we obtain cache-optimal algorithms for some fundamental problems like
sorting, FFT, and an important subclass of permutations in the
single-level cache model.
We also show that ignoring associativity concerns could lead to
inferior performance,
by analyzing the average-case cache behavior of mergesort.
We further extend our model to multiple levels of cache with limited
associativity and present optimal algorithms for matrix transpose and sorting.
Our techniques may be used for systematic exploitation of the memory
hierarchy starting from the algorithm design stage,
and dealing with the hitherto unresolved problem of limited
associativity.
\end{abstract}


\section{Introduction}
\LABEL{sec}{intro}

Models of computation are essential for abstracting the complexity of
real machines and enabling the design and analysis of algorithms.
The widely-used RAM model owes its longevity and usefulness to its
simplicity and robustness.
While it is far removed from the complexities of any physical
computing device,
it successfully predicts the relative performance of algorithms based
on an abstract notion of operation counts.

The RAM model assumes a flat memory address space with unit-cost
access to any memory location.
With the increasing use of caches in modern machines,
this assumption grows less justifiable.
On modern computers,
the running time of a program is as much a function of operation
count as of its cache reference pattern.
A result of this growing divergence between model and reality is that
operation count alone is not always a true predictor of the running
time of a program,
and manifests itself in anomalies such as a matrix multiplication
algorithm demonstrating $O(n^5)$ running time instead of the expected
$O(n^3)$ behavior predicted by the RAM model~\cite{ACFS:94}.
Such shortcomings of the RAM model motivate us to seek an alternative
model that more realistically models the presence of a memory
hierarchy.
In this paper,
we address the issue of better and systematic utilization of caches
starting from the algorithm design stage.

A challenge in coming up with a good model is achieving a balance
between abstraction and fidelity,
so as not to make the model unwieldy for theoretical analysis or
simplistic to the point of lack of predictiveness.
Memory hierarchy models used by computer architects to design caches
have numerous parameters and suffer from the first
shortcoming~\cite{aah,przybylski90}.
Early algorithmic work in this area focussed on a two-layered memory
model\cite{HK:81}---a very large capacity memory with slow access time
(secondary memory) 
and a limited size faster memory (internal memory). All computation is
performed on elements in the internal memory and there is no restriction
on placement of elements in the internal memory (fully associative).

The focus of this paper is on the interaction between main memory and
\emph{cache}, which is the first level of memory hierarchy once the
address is provided by the CPU.
The structure of 
a single level hierarchy of cache memory
is adequately characterized by the following three parameters:
\textbf{A}ssociativity,
\textbf{B}lock size,
and \textbf{C}apacity.
Capacity and block size are in units of the minimum memory access size
(usually one byte).
A cache can hold a maximum of $C$ bytes.
However,
due to physical constraints,
the cache is divided into \emph{cache frames} of size $B$ that contain
$B$ contiguous bytes of memory---called a \emph{memory block}.
The associativity $A$ specifies the number of different frames in
which a memory block can reside.
If a block can reside in any frame
(i.e., $A = \frac{C}{B}$),
the cache is said to be \emph{fully associative};
if $A = 1$, 
the cache is said to be \emph{direct-mapped};
otherwise,
the cache is \emph{$A$-way set associative}.

For a given memory access,
the hardware inspects the cache to determine if the corresponding
memory element is resident in the cache.
This is accomplished by using an indexing function to locate the
appropriate set of cache frames that may contain the memory block.
If the memory block is not resident,
a cache miss is said to occur. 
From an architectural standpoint,
cache misses can be classified into one of three
classes~\cite{hill:evaluating}.
\begin{itemize}
\item
A \emph{compulsory miss} 
(also called a \emph{cold miss})
is one that is caused by referencing a
previously unreferenced memory block.
Eliminating a compulsory miss requires prefetching the data, 
either by an explicit prefetch operation or by placing more data items
in a single memory block. 
\item
A reference that is not a compulsory miss but misses in a
fully-associative cache with LRU replacement is classified as a
\emph{capacity miss}. 
Capacity misses are caused by referencing more memory blocks than can
fit in the cache.
Restructuring the program to re-use blocks while they are in cache can
reduce capacity misses.
\item
A reference that is not a compulsory miss that hits in a
fully-associative cache but misses in an $A$-way set-associative cache
is classified as a \emph{conflict miss}.
A conflict miss to block X indicates that block X has been referenced
in the recent past,
since it is contained in the fully-associative cache, 
but at least $A$ other memory blocks that map to the same cache set
have been accessed since the last reference to block X.
Eliminating conflict misses requires transforming the program to
change either the memory allocation and/or layout of the two arrays
(so that contemporaneous accesses do not compete for the same sets)
or the manner in which the arrays are accessed.
\end{itemize}
Conflict misses pose an additional challenge in designing efficient
algorithms in the cache.
This class of misses is not present in the I/O models,
where the mapping between internal and external memory is fully
associative.

Existing memory hierarchy models~\cite{AV:88,AACS:87,ACS:87,ACFS:94}
do not model certain salient features of caches,
notably the lack of full associativity in address mapping and the lack
of explicit control over data movement and replacement. 
Unfortunately, these small differences are malign in the effect.\footnote{See
the discussion in \cite{CG:98} on a simple matrix transpose program.}
The \emph{conflict misses} that they introduce
make analysis of algorithms
much more difficult~\cite{fricker95}.
Carter and Gatlin~\cite{CG:98} conclude a recent paper saying
\begin{quote}
 \emph{What is needed next is a study of ``messy details'' not modeled
       by UMH (particularly cache associativity) that are important to
       the performance of the remaining steps of the FFT algorithm.}
\end{quote} 

In the first part of this paper,
we develop a two-level memory hierarchy model to capture the
interaction between cache and main memory.
Our model is a simple extension of the two-level I/O model that Aggarwal
and Vitter~\cite{AV:88} proposed for analyzing external memory
algorithms.
However,
it captures three additional constraints of caches:
lower miss penalties;
lack of full associativity in address mapping;
and lack of explicit program control over data movement.
The work in this paper shows that the constraint imposed by limited
associativity can be tackled quite elegantly,
allowing us to extend the results of the I/O model to the cache model
very efficiently.

Most modern architectures
have a memory hierarchy consisting of multiple cache levels.
In the second half of this paper,
we extend the two-level cache model to a multi-level cache model.

The remainder of this paper is organized as follows.
\secref{related} surveys related work.
\secref{emulation} defines our cache model and establishes an
efficient emulation scheme between the I/O model and our cache model.
As direct corollaries of the emulation scheme,
we obtain cache-optimal algorithms for several fundamental problems
such as sorting, FFT, and an important class of permutations.
\secref{avg-case-mergesort} illustrates the importance of the
emulation scheme by demonstrating that a direct
(\ie, bypassing the emulation)
implementation of an I/O-optimal sorting algorithm
(multiway mergesort) is provably inferior,
even in the average case, in the cache model.
\secref{multi-level} describes a natural extension of our model to
multiple levels of caches.
We present an algorithm for transposing a matrix in the
multi-level cache model that attains optimal performance in the presence
of any number of levels of cache memory. Our algorithm is not
cache-oblivious, \ie,
we do make explicit use of the sizes of the cache at various levels.
Next, we show that with some simple modifications, the funnel-sort algorithm
of Frigo et al. attains optimal performance in a single level
(direct mapped) cache in an oblivious sense, \ie,
without prior knowledge of memory parameters. 
Finally,
\secref{concl} presents conclusions,
possible refinements to the model,
and directions for future work.


\section{Related work}
\LABEL{sec}{related}

The I/O model assumes that most of the data resides on disk and has to
be transferred to main memory to do any processing.
Because of the tremendous difference in speeds, 
it ignores the cost of internal processing and counts only the number
of I/Os.  
Floyd~\cite{F:72} originally defined a formal model and proved tight
bounds on 
the number of I/Os required to transpose a matrix using two pages of
internal memory.
Hong and Kung~\cite{HK:81} extended this model and studied the I/O
complexity of FFT when the internal memory size is bounded by $M$.
Aggarwal and Vitter~\cite{AV:88} further refined the model by
incorporating an additional parameter $B$,
the number of (contiguous) elements transferred in a single I/O
operation.
They gave upper and lower bounds on the number of I/Os for several
fundamental problems including sorting, 
selection, 
matrix transposition,
and FFT.
Following their work,
researchers have designed I/O-optimal algorithms for fundamental
problems in graph theory~\cite{CGG:95} and computational
geometry~\cite{GTV:93}.

Researchers have also modeled multiple levels of memory hierarchy.
Aggarwal \etal~\cite{AACS:87} defined the \emph{Hierarchical Memory
Model} 
(HMM)
that assigns a function $f(x)$ to accessing location $x$ in the
memory,
where $f$ is a monotonically increasing function. 
This can be regarded as a continuous analog of the multi-level
hierarchy. 
Aggarwal \etal~\cite{ACS:87} added the capability of block transfer
to the HMM,
which enabled them to obtain faster algorithms.
Alpern \etal~\cite{ACFS:94} described the 
\emph{Uniform Memory Hierarchy} (UMH) model, 
where the access costs differ in discrete steps. 
Very recently, Frigo \etal~\cite{FLPR:99} presented an alternate strategy
of algorithm design on these models which has the added advantage
that explicit values of parameters related to different  
levels of the memory hierarchy are not required. Bilardi and Peserico
\cite{BP:00} investigate further the complexity of designing 
algorithms without the knowledge architectural 
parameters.\footnote{However,
none of these models address the problem of limited associativity in
cache.}
Other attempts were directed towards extracting better performance by
parallel memory hierarchies~\cite{NV:91,VS:94,cormen99}, 
where several blocks could be transferred simultaneously.
 
Ladner \etal~\cite{LFL:99} describe a stochastic model for performance
analysis in cache.
Our work is different in nature,
as we follow a more traditional worst-case analysis. 
Our analysis of sorting in \secref{avg-case-mergesort} provides a
better theoretical basis for some of the experimental work of LaMarca
and Ladner~\cite{LL:97}.

To the best of our knowledge, the only other paper that addresses the 
problem of limited associativity in cache is 
recent work of Mehlhorn and Sanders\cite{Sanders:99}. 
They show that for a class
of algorithms based on merging multiple sequences, the I/O algorithms
can be made nearly optimal by use of a simple randomized shift technique.
The emulation theorem in \secref{emulation} of this 
paper not only provides a deterministic solution for the same class of
algorithms, but also works for a very general situation. The results in 
\cite{Sanders:99} are nevertheless
interesting from the perspective of implementation.


\def\buf{\mbox{\emph{Buf}}\xspace}
\def\mem{\mbox{\emph{Mem}}\xspace}
\def\io{\ensuremath{{\mathfrak{I}}}}
\def\cache{\ensuremath{\mathfrak{C}}}

\section{The cache model}
\label{secemu}
\LABEL{sec}{emulation}

The (two-level) I/O model of Aggarwal and Vitter~\cite{AV:88} captures
the interaction between a slow (secondary) memory of infinite capacity
and a fast (primary) memory of limited capacity.
It is characterized by two parameters:
$M$, the capacity of the fast memory; and
$B$, the size of data transfers between slow and fast memories.
Such data movement operations are called \emph{I/O operations} or
\emph{block transfers}.
The use of the model is meaningful when the problem size $N \gg M$.

The I/O model contains the following further assumptions.
\begin{enumerate}
\item
A datum can be used in a computation iff it is present in fast
memory.
All data initially resides in slow memory.
Data can be transferred between slow and fast memory
(in either direction)
by I/O operations.

\item
Since the latency for accessing slow memory is very high,
the average cost of transfer per element can be reduced by
transferring a block of $B$ elements at little additional cost.
This may not be as useful as it may seem at first sight,
since these $B$ elements are not arbitrary,
but are contiguous in memory.
The onus is on the programmer to use all the elements,
as traditional RAM algorithms are not designed for such restricted
memory access patterns.
We denote the map from a memory address to its block address by \BB.
The internal memory can hold at least three blocks, i.e., 
$M \geq 3\cdot B$.

\item
The computation cost is ignored in comparison to the cost of an I/O
operation.
This is justified by the high access latency of slow memory.

\item
A block of data from slow memory can be placed in any block of fast
memory.

\item
I/O operations are explicit in the algorithm.
\end{enumerate}
The goal of algorithm design in this model is to minimize the number
of I/O operations.

We adopt much of the framework of the I/O model in developing a cache
model to capture the interactions between cache and main memory.
In this case,
the cache is the fast memory,
while main memory is the slow memory.
Assumptions 1 and 2 of the I/O model continue to hold in our cache
model.
However,
assumptions 3--5 are no longer valid and need to be replaced as
follows.
\begin{itemize}
\item
The difference between the access times of slow and fast memory is
considerably smaller than in the I/O model,
namely a factor of 5--100 rather than factor of 10000.
We will use $L$ to denoted the \emph{normalized} cache latency.
This cost function assigns a cost of 1 for accessing an element in
cache and $L$ for accessing an element in the main memory.
This way, we also account for the computation in cache. 

\item
Main memory blocks are mapped into cache sets using a \emph{fixed} and
pre-determined mapping function that is implemented in hardware.
Typically,
this is a modulo mapping based on the low-order address bits.
However,
the results of this section will hold as long as there is a
\emph{fixed} address mapping function that distributes the main memory
evenly in the cache.
We denote this mapping from main memory blocks to cache sets by \BS.
We will occasionally slightly abuse this notation and apply \BS\ 
directly to a memory address $x$ rather than to $\BB(x)$.

\item
The cache is not visible to the programmer 
(not even at the assembly level). 
When a program issues a reference to a memory location $x$,
an \emph{image} (copy) of the main memory block $b = \BB(x)$ is
brought into the cache set $\BS(b)$ if it is not already present
there.
The block $b$ continues to reside in cache until it is evicted by
another block $b^\prime$ that is mapped to the same cache set
(\ie, $\BS(b) = \BS(b^\prime)$).
In other words,
a cache set $c$ contains the latest memory block referenced that is
mapped to this set.
\end{itemize}

To summarize,
we use the notation  $\cache (M , B, L )$ to  denote our
three-parameter cache model,
and the notation $\io (M, B)$ to denote the I/O model with parameters
$M$ and $B$.
We will use $n$ and $m$ to denote
$N/B$ and $M/B$ respectively. 
The assumptions of our cache model parallel those of the I/O model,
except as noted above.\footnote{Frigo \etal~\cite{FLPR:99}
independently arrive at a very similar parameterization of their
model.}
The goal of algorithm design in the cache model is to minimize
\emph{running time},
defined as the number of cache accesses plus $L$ times the number of
main memory accesses.

\subsection{Emulating I/O algorithms}
  
The differences between the two models listed above would appear to
frustrate any efforts to naively map an I/O algorithm to the cache
model,
given that we neither have the control nor the flexibility of the I/O
model.
Our main result in this section establishes a connection between the
I/O model and the cache model using a very simple emulation scheme.

\begin{theorem}[Emulation Theorem]
An algorithm $A$ in $\io (M, B)$ using $T$ block transfers and $I$
processing time can be converted to an equivalent algorithm $A^c$ in
$\cache (M, B, L)$ that runs in $O( I + (L + B)\cdot T)$ steps.
The memory requirement of $A^c$ is an additional $m +2$ blocks beyond
that of $A$.
\label{u_bnd}
\LABEL{thm}{u_bnd}
\end{theorem} 

\begin{proof}
Note that $I$ is usually not accounted for in the I/O model,
but we will keep track of the internal memory computation done in $A$
in our emulation.
The idea behind the emulation is as follows. 
We will mimic the behavior of the I/O algorithm $A$ in the cache
model,
using an array \buf of $m$ blocks to play the role of the fast
memory. 
We will view the main memory in the cache model as an array \mem of
$B$-element blocks.
Although \buf is also part of the memory, 
we are using different notations to make their roles explicit in
this proof. 
Likewise,
we will view the cache as an array of sets and denote the $i$th set by
$C[i]$.

As discussed above, 
we do not have explicit control on the contents of the cache
locations. However, 
we can control the memory access pattern through a level of
indirection so as to maintain a 1-1 correspondence between \buf and
the cache. 
Wlog, 
we assume that \BS\ maps block $i$ of \buf to cache set $C[i]$ for $i
\in [1, m]$.

We divide the I/O algorithm into rounds, 
where in each round, 
the I/O algorithm $A$ transfers a block between the slow memory and
the fast memory and (possibly) does some computations.  
The cache algorithm $A^c$ transfers the same blocks between \mem and
\buf and then does the identical computations in \buf.
Figure \ref{emproc} formally describes the procedure. 
Note that the $B$ elements must be explicitly copied in the cache
model.

\begin{figure}
\begin{center}
\begin{tabular}{|l|l|}
\multicolumn{2}{c}{\textbf{Round $t$ of the emulation}} \\ \hline
\emph{I/O Algorithm $A$} & \emph{Cache Emulation $A^c$} \\ \hline
\parbox{2.8in}{1. Transfer block $b_t$ from slow memory to block $a_t$
of the fast memory} 
&
\parbox{2.8in}{ 1. Copy contents of the $B$ locations of\\
 \hspace*{0.2in} $\mem [ b_t ]$ into $\buf[ a_t ] $ 
}
\\
\parbox{2.8in}{2. Perform computations in fast memory} &
\parbox{2.8in}{2. Perform identical computations in \buf} \\
\hline
\end{tabular}
\end{center}
\caption{The emulation scheme used in the proof of \thmref{u_bnd}.}
\label{emproc}
\end{figure}

It must be obvious that the final outcome of algorithm $A^c$ is the
same as algorithm $A$.
The more interesting issue is the cost of the emulation.

A block of size $B$ is transferred into cache if its image does not
exist in the cache at the time of reference. 
The invariant that we try to maintain at the end of each round is that
there is a 1-1 correspondence between \buf and $C$.  
This will ensure that all the $I$ operations are done within the cache
at minimal cost.

Assume that we have maintained the above invariant at the end of round
$t-1$.  
In round $t$, 
we transfer block $\mem [ b_t ]$ into $\buf[ a_t ]$. 
Accessing the memory block $\mem [ b_t ]$ will displace the existing
block in cache set $C[q]$, 
where $q = \BS( b_t )$.
From the invariant, 
we know that the block displaced from $C[q]$ is $\buf[q]$,
which must be restored to cache to restore the invariant.
We can bring it back by a single memory reference and charge this to
the round $t$ itself, 
which is $L$.  
(Actually it will be brought back during the subsequent reference, 
so the previous step is only to simplify the accounting.)

The cost of copying $\mem [ b_t ]$ to $\buf[ a_t ]$ is $L + B$ assuming
that $\mem [ b_t ]$ and $\buf [ a_t ]$ are not mapped to the same cache
set
($\BS(b_t) \neq \BS(a_t)$).  
Otherwise it will cause alternate cache misses (\emph{thrashing}) 
of the blocks $\mem [ b_t ]$ and
$\buf [ a_t ]$ leading to $L\cdot B$ steps for copying. 
This can be prevented by transferring through an intermediate memory
block $\mem [ Y ]$ such that $\BS (Y) \neq \BS ( b_t )$.  
Having two such intermediate buffers that map to distinct cache sets
would suffice in all cases.  
So, 
we first transfer $\mem [ b_t ]$ to $\mem [ Y ]$ followed by $\mem [ Y ]$ to
$\buf[ j ]$. 
The first copying has cost $2L +B$ since both blocks must be fetched
from main memory.  
The second transfer is between blocks, 
one of which is present in the cache, 
so it has cost $L+B$.  
To this we must also add cost $L$ for restoring the block of \buf
that was mapped to the same cache set as $\mem [ Y ]$. 
So, 
the total cost of the \emph{safe} method is $4L + 2B$.

The internal processing remains identical. 
If $I_t$ denotes the internal processing cost of step $t$, 
the total cost of the emulation is at most $\sum_{t=1}^{T} ( I_t +
2(L + B) + 2L ) = I + 4L \cdot T + 2B\cdot T$.
\end{proof}

\begin{remark}
\end{remark}
\begin{itemize}
\item 
A possible alternative to using intermediate memory-resident buffers
to avoid thrashing is to use registers, 
since register access is much faster. 
In particular, 
if we have $B$ registers, 
then we can save two extra memory accesses, 
bringing down the emulation cost to $2L + 2B$.
\item We can make the emulation somewhat simpler by using a randomized
mapping scheme. That is, 
if we choose the starting location of array \buf randomly, 
then the probability that $\mem [ b_t ]$ and $\buf[ a_t ]$ have the same
image is $1/M$. 
So the expected emulation cost is $I + 2L \cdot T + (B + (LB)/M) \cdot
T $ without using any intermediate copying.
\item The basic idea of copying data into contiguous memory locations to 
reduce interference misses has been exploited before
in some specific contexts like matrix
multiplication~\cite{lam:blocked} and bit-reversal
permutation~\cite{CG:98}.
\thmref{u_bnd} unifies these previous results within a common
framework.
\end{itemize}

The term $O(B\cdot T)$ is subsumed by $O(I)$ if computation is done on
at least a constant fraction of the elements in the block transferred
by the I/O algorithm. 
This is usually the case for efficient I/O algorithms. 
We will call such I/O algorithms \emph{block-efficient}.

\begin{corollary}
A block-efficient I/O algorithm for $\io (M, B)$ 
that uses $T$ block transfers and $I$
processing can be emulated in $\cache (M, B, L)$ in $O( I + L\cdot T)$
steps.
\label{goodemu}
\end{corollary}
\begin{remark}
\LABEL{remark}{blkeff}
The algorithms for sorting, 
FFT, 
matrix transposition, 
and matrix multiplication described in Aggarwal and
Vitter~\cite{AV:88} are block-efficient.
\end{remark}
  
\subsection{Extension to set-associative cache}

The trend in modern memory architectures is to
allow limited flexibility in the address mapping between memory blocks
and cache sets. 
The $k$-way set-associative cache has the property that a memory block
can reside in any (one) of $k$ cache frames. 
Thus,
$k =1$ corresponds to the direct-mapped cache we have considered so
far,
while $k=m$ corresponds to a fully associative cache. 
Values of $k$ for data caches are generally small, usually in the
range 1--4.

If all the $k$ sets are occupied, 
a replacement policy like LRU is used 
(by the hardware) 
to find an assignment for the referenced block. 
The emulation technique of the previous section would extend to this
scenario easily if we had explicit control on the replacement. 
This not being the case, 
we shall tackle it indirectly by making use of an useful property of
LRU that Frigo \etal~\cite{FLPR:99} exploited in the context of
designing cache-oblivious algorithms for a fully associative cache.
\begin{lemma}[Sleator-Tarjan\cite{ST:85}]
For any sequence $s$, $F_{LRU}$, the number of misses incurred by LRU with
cache size $n_{LRU}$ is no more than 
$( n_{LRU}/( n_{LRU} - n_{OPT} +1) F_{OPT})$,
where $F_{OPT}$ is the minimimum number of misses by an optimal
replacement strategy with cache size $n_{OPT}$.
\label{ST}
\end{lemma}
We use this lemma in the following way. 
We run the emulation technique for only half the cache size, 
\ie, 
we choose the buffer to be of size $m/2$, 
such that for every $k$ cache lines in a set, 
we have only $k/2$ buffer blocks. 
From Lemma \ref{ST}, 
we know that the number of misses in each each cache set is no more
than twice the optimal,
which is in turn bounded by the number of misses incurred by the I/O
algorithm.
\begin{theorem}[Generalized Emulation Theorem]
An algorithm $A$ in $\io (M/2, B)$ using $T$ block transfers and $I$
processing time can be converted to an equivalent algorithm $A^c$ in
the $k$-way set-associative cache model with parameters $M,B, L$
that runs in $O( I + (L + B)\cdot T)$ steps.
The memory requirement of $A^c$ is an additional $m/2 +2$ blocks beyond
that of $A$.
\end{theorem}


\subsection{The cache complexity of sorting and other problems}
\LABEL{sec}{sorting}

Aggarwal and Vitter~\cite{AV:88} prove the following lower bound
for sorting and FFT in the I/O model.
\begin{lemma}[\cite{AV:88}]
The average-case and the worst-case number of I/O's required for
sorting $N$ records and for computing the $N$-input FFT graph
in $\io (M, B)$ is $\Omega \left( \frac{N}{B}\,\frac{\log (1 + N/B)}{\log (1 + M/B )}
\right)$.
\label{srtlb}
\end{lemma}

\begin{theorem}
The lower bound for sorting in $\cache (M, B, L)$ is 
$\Omega ( N\log N + L \frac{N}{B}\,\frac{\log {N/B}}{\log {M/B}})$.
\label{lbnd_srt}
\end{theorem}

\begin{proof}
Any lower bound in the number of block transfers in $\io (M, B)$
carries over to $\cache (M, B, L)$.
Since the lower bound is the maximum of the lower bound on 
number of comparisons and the bound in Lemma \ref{srtlb}, 
the theorem follows by dividing the sum of the two terms by 2. 
\end{proof}

\begin{theorem}
In $\cache (M, B, L)$, $N$ numbers can be sorted in $O( N\log N + 
 L \cdot \frac{N}{B} \cdot \frac{\log {N/B}}{\log {M/B}}) $ steps and
this is optimal.
\end{theorem}

\begin{proof}
The $M/B$-way mergesort algorithm described in
Aggarwal and Vitter \cite{AV:88} has an I/O complexity of 
$O(\frac{N}{B} \frac{\log {N/B}}{\log {M/B}} )$. 
The processing time involves maintaining a heap of size $M/B$ and
$O(\log M/B )$ per  output element. 
For $N$ elements, 
the number of phases is $\frac{\log N}{\log M/B}$, 
so the total processing time is $O(N\log N)$.
From Corollary \ref{goodemu}, 
and Remark~\ref{remark:blkeff},
the cost of this algorithm in the cache model
is $O( N\log N + L \cdot \frac{N}{B} \cdot \frac{\log {N/B}}{\log
{M/B}}) $.
Optimality follows from Theorem \ref{lbnd_srt}.
\end{proof}

\begin{remark}
The $M/B$-way distribution sort (multiway quicksort) also has the same
upper bound.
\end{remark}

We can prove a similar result for FFT computations.
\begin{theorem}
The FFT of $N$ numbers can be computed in $O(
N\log N +  L \cdot \frac{N \log {N/B}}{B \log {M/B}}) $ in $\cache (M, B, L)$. 
\end{theorem}

\begin{remark}
The FFTW algorithm \cite{frigo98} is optimal only for $B = 1$. 
Barve \cite{B:99} has independently obtained a similar result.
\end{remark}
The class of Bit Matrix Multiply Complement (BMMC) permutations 
include many important permutations like matrix transposition and
bit reversal. 
Combining the work of Cormen \etal~\cite{cormen99} with our emulation
scheme,
we obtain the following result.

\begin{theorem}
The class of BMMC permutations for $N$ elements can be achieved in
$\Theta \left( N + L\cdot \frac{N}{B} \frac{\log M}{\log (M/B)}
\right)$ steps in $\cache( M, B, L)$.
\label{bmmc}
\end{theorem}
\begin{remark}
Many known geometric~\cite{CGG:95} and graph algorithms~\cite{GTV:93}
in the I/O model,
such as convex hull and graph connectivity,
can be transformed optimally into the cache model. 
\end{remark}


\section{Average-case performance of mergesort in the cache model}
\LABEL{sec}{avg-case-mergesort}

In this section, 
we analyze the average-case performance of $k$-way mergesort in the
cache model.
Of the three classes of misses described in \secref{intro},
we note that compulsory misses are unavoidable and that capacity
misses are minimized while designing algorithms for the I/O model.
We are therefore interested in bounding the number of conflict
misses for a straightforward implementation of the I/O-optimal $k$-way
mergesort algorithm.
It is easy to construct a worst-case input permutation where there
will be a conflict miss for every input element 
(a cyclic distribution suffices), 
so the average case is more interesting.

We assume that $s$ cache sets are available for the leading blocks of
the $k$ runs $S_1, \ldots , S_k$.
In other words, 
we ignore the misses caused by heap operations 
(or equivalently ensure that the heap area in the cache does not
overlap with the runs). 

We create a random instance of the input as follows.
Consider the sequence $\{1, \ldots , N\}$,
and distribute the elements of this sequence to runs by traversing
the sequence in increasing order and assigning element $i$ to run
$S_j$ with probability $1/k$.
From the nature of our construction, 
each run $S_i$ is sorted. 
We denote $j$-th element of $S_i$ as $S_{i,j}$.
The expected number of elements in any run $S_i$ is $N/k$.

During the $k$-way merge, 
the leading blocks are critical in the sense that the heap is built on
the \emph{leading element} of every sequence $S_i$.  
The leading element of a sequence is the smallest element that has not
been added to the merged (output) sequence. 
The \emph{leading block} is the cache line containing the leading
element.  
Let $b_i$ denote the leading block of run $S_i$.  
\emph{Conflict} can occur when the leading blocks of different
sequences are mapped to the same cache set. 
In particular, 
a \emph{conflict miss} occurs for element $S_{i,j+1}$ when there
is at least one element $x \in b_k$, 
for some $ k \neq i$,
such that $S_{i,j} < x < S_{i,j+1}$ and $\BS ( b_i ) = \BS ( b_k )$. 
(We do not count conflict misses for the first element in the
leading block, 
\ie, $S_{i,j}$ and $S_{i,j+1}$ must belong to the same block,
but we will not be very strict about this in our calculations.)

Let $p_i$ denote the probability of conflict for element $i
\in [1, N]$.
Using indicator random variables $X_i$ to count the conflict miss for
element $i$, 
the total number of conflict misses $X = \sum_i X_i$. 
It follows that the expected number of conflict misses $E [ X ] =
\sum_i E [ X_i ] = \sum_i p_i$. 
In the remaining section we will try to estimate a lower bound on
$p_i$ for $i$ large enough to validate the following assumption.
\begin{quote}
{\bf A1} 
The cache sets of the leading blocks, 
$\BS ( b_i )$,
are randomly distributed in cache sets $1, \ldots, s$
independent of the other sorted runs.
Moreover, 
the exact position of the leading element within the leading block is
also uniformly distributed in positions $\{ 1, \ldots, sB \}$. 
\end{quote}

\begin{remark} 
A recent variation of the mergesort algorithm
(see~\cite{BGV:96}) actually satisfies {\bf A1} by its very nature.
So, the present analysis is directly applicable to its average-case
performance in cache. A similar observation was made independently by
Sanders~\cite{Sanders:99} who
obtained upper-bounds for mergesort for a set associative cache.
\end{remark}

From our previous discussion and the definition of a conflict
miss, 
we would like to compute the probability of the following event.
\begin{quote}
{\bf E1} For some $i,j$, 
for all elements $x$, 
such that $S_{i,j} < x < S_{i,j+1}$, $\BS (x) \neq \BS ( S_{i,j})$.
\end{quote}
In other words, 
none of the leading blocks of the sorted sequences $S_j$, $j \neq i$,
conflicts with $b_i$. 
The probability of the complement of this event
(\ie, $\Pr [ \overline{E1} ]$)
is the probability that we want to estimate.
We will compute an upper bound on $\Pr [ E1 ]$,
under the assumption A1, 
thus deriving a lower bound on $\Pr[ \overline{E1} ]$.

\begin{lemma}
For $k/s > \epsilon$, 
$\Pr[ E1 ] < 1 - \delta$, 
where $\epsilon$ and $\delta$ are positive constants
(dependent only on $s$ and $k$ but not on $n$ or $B$).  
\label{weakbnd}
\end{lemma}  

\begin{proof}
See \appref{approx}.
\end{proof}

Thus we can state the main result of this section as follows.
\begin{theorem}
The expected number of conflict misses in a random input for doing
a $k$-way merge in an $s$-set direct-mapped cache, 
where $k$ is $\Omega (s)$, is $\Omega (N)$,
where $N$ is the total number of elements in all the $k$
sequences.
Therefore the 
(ordinary I/O-optimal) 
$M/B$-way mergesort in an $M/B$-set cache will exhibit $O(N \frac{\log
N/B}{\log M/B})$ cache misses which is asymptotically larger than the
optimal value of $O(\frac{N}{B} \frac{\log N/B}{\log M/B})$.
\end{theorem}

\begin{proof}
The probability of conflict misses is $\Omega (1)$ when $k$ is
$\Omega (s)$. 
Therefore the expected total number of conflict misses is $\Omega
(N)$ for $N$ elements. 
The I/O-optimal mergesort uses $M/B$-way merging at each of the $
\frac{\log N/B}{\log M/B} $ levels, 
hence the second part of the theorem follows.
\end{proof}

\begin{remark}
Intuitively, by choosing $k \ll s$, 
we can minimize the probability of conflict misses 
resulting in an increased number of merge phases (and hence
running time).
This underlines the critical role of conflict misses {\em vis-a-vis}
capacity misses that 
forces us to use only a small fraction of the available cache. 
Recently, Sanders~\cite{Sanders:99} has shown that by choosing $k$ to be
$O(\frac{M}{B^{1+ 1/a}})$ in an $a$-way set associative cache 
with a modified version of mergesort of \cite{BGV:96}, the
expected number of conflict misses per phase can be bounded by $O(N/B)$.
In comparision, the use of the emulation theorem guarantees minimal 
worst-case conflict misses while making good use of cache.
%
\end{remark}


\section{The Multi-level Cache Model}
\LABEL{sec}{multi-level}

Most modern architectures
have a memory hierarchy consisting of multiple levels of cache.
Consider two cache levels $\mathcal{L}_1$ and $\mathcal{L}_2$ preceding 
main memory,
with $\mathcal{L}_1$ being faster and smaller.
The operation of the memory hierarchy in this case is as follows.
The memory location being referenced is first looked up in $\mathcal{L}_1$.
If it is not present in $\mathcal{L}_1$,
then it is searched for in $\mathcal{L}_2$
(these can be overlapped with appropriate hardware support).
If the item is not present in $\mathcal{L}_1$ but it is in $\mathcal{L}_2$,
then it is brought into $\mathcal{L}_1$.
In case that it is not in $\mathcal{L}_2$,
then a cache line is brought in from main memory into $\mathcal{L}_2$
and into $\mathcal{L}_1$.
The size of cache line brought into $\mathcal{L}_2$
(denoted by $B_2$)
is usually no smaller than the one brought into $\mathcal{L}_1$
(denoted by $B_1$).
The expectation is that the more frequently used items will remain in
the faster cache.

The Multi-level Cache Model is an extension to multiple cache levels of the 
previously introduced Cache Model. 
Let  $\mathcal{L}_i$ denote the $i$-th level of cache memory.
The parameters involved here are the problem size {\em N}, 
the size of $\mathcal{L}_i$ which is denoted by 
{\em $M_i$}, the frame size (unit of allocation)
of $\mathcal{L}_i$ denoted by $B_i$ and the latency factor $l_i$. 
If a data item is present in the $\mathcal{L}_i$, then it is present in
$\mathcal{L}_j$ for all $j \geq i$ (sometimes referred to as the {\bf
inclusion property}). If it is not present in $\mathcal{L}_i$,
then the cost for a miss is $l_i$ plus the cost of 
fetching it from $\mathcal{L}_{i+1}$ (if
it is present in $\mathcal{L}_{i+1}$, then this cost is zero).
For convenience, the latency factor $l_i$ 
is the ratio of time taken on a miss from the $i$-th level 
to the amount of time taken for a unit operation. 

\begin{figure}[tbp]
\begin{center}
\epsfig{file=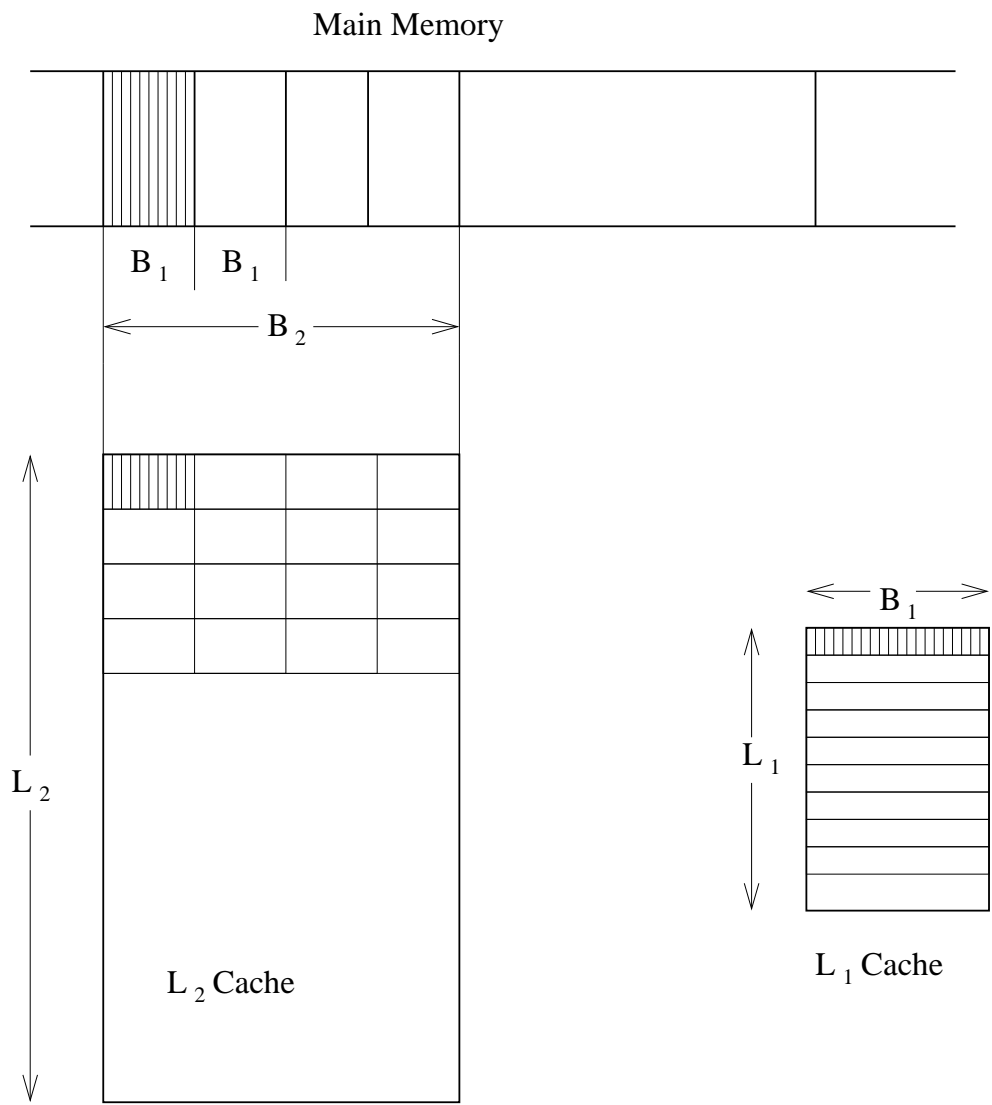,width=\linewidth}
\end{center}
\caption{Memory mapping in a two-level cache hierarchy}
\label{2lcache}
\end{figure}

Figure \ref{2lcache} shows the memory mapping for a two-level cache 
architecture. 
The shaded part of main memory is of size $B_1$ and therefore occupies
only a part of a line of the $\mathcal{L}_2$ cache which is of size
$B_2$. 
There is a natural generalization of the memory mapping to multiple
levels of cache.

We make the following assumptions in this section,
which are consistent with existing architectures.
\begin{quote}
A1. For all $i$, $B_i$, $L_i$ are powers of 2.\\
A2. $2 B_i \leq B_{i+1}$ and the 
number of Cache Lines $L_i \le L_{i+1}$.\\
A3. 
$B_k \le L_1$ and $4 B_k \le B_1 L_1$ (i.e. $B_1 \geq 4$) 
where $\mathcal{L}_k$ is the largest and 
slowest cache. This implies that 
\begin{equation}
L_i \cdot B_i \geq B_k \cdot B_i
\label{bandsize}
\end{equation}
This will be useful for the analysis of the algorithms and are sometimes
termed as {\em tall cache} in reference to the aspect ratio.
\end{quote}

\subsection{Matrix Transpose}

In this section, 
we provide an approach for transposing a matrix in the Multi-level
Cache Model.

The trivial lower bound for matrix transposition of an $N \times N$
matrix in the multi-level cache hierarchy is clearly the time to scan
$N^2$ elements, namely,
\[	
\Omega (\sum_i \frac{N^2}{B_i}l_i )
\]
where \\
$B_i$ is the number of elements in one cache line in $\mathcal{L}_i$
cache;
$L_i$ is the number of cache lines in $\mathcal{L}_i$ cache, 
which is $\frac{M_i}{B_i}$;
and $l_i$ is the latency for $\mathcal{L}_i$ cache.

Our algorithm uses a more general form of the emulation theorem to get
the submatrices to fit into cache in a regular fashion.
The work in this section shows that it is possible to handle the
constraints imposed by limited associativity even in a multi-level
cache model.

	We subdivide the matrix into $B_k\times B_k$ submatrices. 
Thus we get $\lceil n/B_k \rceil \times \lceil n/B_k \rceil$ 
submatrices from an $n \times n$ submatrix.


\[
A = 
\left(
\begin{array}{ccccc}
a_1	       & a_2           & \ldots & \ldots 	& a_n    \\
a_{n+1}	       & a_{n+2}       & \ldots & \ldots 	& a_{2n} \\
\vdots         & \vdots	       & \vdots	& \vdots 	& \vdots \\
\vdots         & \vdots	       & \vdots	& \vdots 	& \vdots \\
a_{n^2-n+1}    & \ldots	       & \ldots & \ldots 	& a_{n^2}
\end{array} 
\right) = 
\left(
\begin{array}{ccccc}
A_1	       & A_2           & \ldots  	& A_{n/B}    \\
\vdots         & \vdots	       & \vdots 	& \vdots \\
A_{n^2-nB/B}   & \ldots	       & \ldots 	& A_{n^2/B^2}
\end{array} 
\right) 
\]

Note that the submatrices in the last row and column need not be square as one side may have $\le B$ rows or columns.

Let $m = n/B$ then 

\[
A^T = 
\left(
\begin{array}{ccccc}
A_1^T	       	& A_{m+1}^T     & \ldots & \ldots 	& A_{m^2-m+1}^T    \\
A_2^T		& A_{m+2}^T     & \ldots & \ldots 	& A_{2m}^T \\
\vdots         	& \vdots	& \vdots & \vdots 	& \vdots \\
\vdots         	& \vdots	& \vdots & \vdots 	& \vdots \\
A_m^T	  	& \ldots	& \ldots & \ldots 	& A_{m^2}^T
\end{array} 
\right)
\]

For simplicity, we describe the algorithm as transposing a
square matrix $A$ in another matrix $B$, i.e. $B = A^T$. The main
procedure is {\bf Rec\_Trans}($A, B, s$), where $A$ is transposed into
$B$ by dividing $A$  and $B$ into $s^2$ submatrices and then recursively
transposing the sub-matrices. Let $A_{i,j}$ ($B_{i,j}$) denote the submatrices
for $1 \leq i,j \leq s$. Then $B = A^T$ can be computed as {\bf Rec\_Trans}(
$A_{i,j} , B_{j,i} , s'$) for all $i,j$ and some appropriate $s'$ which
depends on $B_k$ and $B_{k-1}$. 
In general, if $t_k , t_{k-1}, \ldots, t_1$ denote the values of $s'$ at the 
$1, 2 \ldots $ level of recursion, 
then $t_i = B_{i+1} / B_i $.  If the submatrices are
 $B_1 \times B_1$ (base case), then perform the transpose exchange of the 
symmetric submatrices directly.
We perform matrix transpose as follows, which is similar to the familiar
recursive transpose algorithm.
\\ 
1. Subdivide the matrix as shown into $B_k \times B_k$ submatrices.\\
2. Move the symmetric submatrices to contiguous memory locations.\\
3. {\bf Rec\_Trans}( $A_{i,j} , B_{j,i} , B_k / B_{k-1}$). \\ 
4. Write back the $B_k \times B_k$ submatrices to original locations.

In the following subsections we analyze the data movement of this algorithm
to bound the number of cache misses at various levels.
\subsection{Moving a submatrix to contiguous locations}
	
To move a submatrix we will move it cache line by cache line. By choice of size of submatrices ($B_k \times B_k$) each row will be an array of size $B_k$, but the rows themselves may be far apart.

\begin{lemma}
If two memory blocks $x$ and $y$ of size $B_k$ are aligned in
$\mathcal{L}_k$-cache map to the same cache set in
$\mathcal{L}_i$-cache for some $1 \le i \le k$,
then $x$ and $y$ map to the same set in each $\mathcal{L}_j$-cache for
all $1 \le j \le i$.
\end{lemma}

\begin{proof}
If $x$ and $y$ map to the same cache set in $\mathcal{L}_i$ cache 
then their $i$-th level memory block numbers 
(to be denoted by $b^i (x)$ and $b^i (y)$)  
differ by a multiple of $L_i$. Let $b^i (x) - b^i (y) = \alpha L_i$. 
Since $L_j | L_i$ (both are powers of two), $b^i (x) - b^i (y) = \beta L_j$
where $\beta = \alpha \cdot L_i / L_j$.
Let $x' , y'$ be the {\em corresponding} sub-blocks of $x$ and $y$ at the 
$j$-th level. Then their block numbers  $b^j ( x' ),  b^j ( y' )$ differ
by $B_i /B_j \cdot \beta \cdot  L_j$, i.e., a multiple of $L_j$ as $B_j | B_i$. 
Note that blocks are aligned across different levels of cache.
%
Therefore $x$ and $y$ also collide in $\mathcal{L}_j$. 
\end{proof}

\begin{corollary}
If two blocks of size $B_k$ that are aligned in $\mathcal{L}_k$-cache 
do not conflict in level $i$ they do not conflict in any level $j$ 
for all $i \le j \le k$.
\end{corollary}

\begin{theorem}
There is an algorithm which moves a set of blocks of size $B_k$ (where
there are $k$ levels of cache with block size $B_i$ for each $1 \le i
\le k$) into a contiguous area in main memory in 
\[
O \left(\sum \frac{N}{B_i}l_i \right)
\]
where N is the total data moved and $l_i$ is the cost of a cache miss
for the $i^{th}$ level of cache.
\label{blkmove}
\end{theorem}

\begin{proof}
	Let the set of blocks of size $B_k$ be $I$ (we are assuming that 
the blocks are aligned). Let the target block in the contiguous area for 
each block $i \in I$ be in the corresponding set $J$ where each block 
$j \in J$ is also aligned with a cache line in $\mathcal{L}_k$ Cache.

	Let block $a$ map to $R_{b,a}$, $b = \{1,2, \ldots ,k\}$ where
$R_{b,a}$ denote the set of cache lines in the $\mathcal{L}_b$-cache.
(Since $a$ is of size $B_k$, it will occupy several blocks in lower
levels of cache.)


Let the $i^{th}$ block map to set $R_{k,i}$  of the $\mathcal{L}_k$
Cache. 
Let the target block $j$ map to set $R_{k,j}$. 
In the worst case,
$R_{k,j}$ is equal to $R_{k,i}$. 
Thus in this case the line $R_{k,i}$ has to be moved to a temporary
block say $x$ (mapped to $R_{k,x}$) and then moved back to $R_{k,j}$.
We choose $x$ such that $R_{1,x}$ and $R_{1,i}$ do not conflict and
also $R_{1,x}$ and $R_{1,j}$ do not conflict. 
Such a choice of $x$ is always possible because our temporary storage
area $X$ of size $4 B_k$ has at least $4$ lines of
$\mathcal{L}_k$-cache ($i$ and $j$ will take up two blocks of
$\mathcal{L}_k$-cache,
thus leaving at least one block free to be used as temporary storage).
{\em This is why we have the assumption that $4 B_k \le B_1 L_1$}. 
That is, 
by dividing the $\mathcal{L}_1$-cache into ${B_1 L_1}/B_k$ zones,
there is always a zone free for $x$.

For convenience of analysis, we maintain the invariant that {\em $X$
is always in $\mathcal{L}_k$-cache}.
By application of the previous corollary on our choice of $x$ 
(such that $R_{1,i} \ne R_{1,x} \ne R_{1,j}$) 
we also have $R_{a,i} \ne R_{a,x} \ne R_{a,j}$ for all $1 \le a \le
k$. 
Thus we can move $i$ to $x$ and $x$ to $j$ without any conflict
misses. 
The number of cache misses involved is three for each level---one for
getting the $i^{th}$ block, 
one for writing the $j^{th}$ block,
and one to maintain the invariant since we have to touch the line
displaced by $i$.
Thus we get a factor of $3$.

Thus the cost of this process is
\[
3 \left(\sum \frac{N}{B_i}l_i \right)
\]
where $N$ is the amount of data moved.

\end{proof}

\begin{remark}
For blocks $I$ that are not aligned in $\mathcal{L}_k$ Cache, the
constant would increase to 4 since we would need to bring up to 2
cache lines for each $i\in I$. The rest of the proof would remain the
same.
\end{remark}

\begin{corollary}
A $B_k \times B_k$ submatrix can be moved into contiguous locations in 
the memory in $O(\sum_{i=1}^{i=k} \frac{ {B_k}^2 }{B_i} l_i )$ time in
a computer that has $k$ levels of (direct-mapped) cache. 
\label{matcopy}
\end{corollary}

This follows from the preceding discussion. We allocate memory say $C$
of size $B_k \times B_k$ for placing the submatrix and memory, say,
$X$ of size $4 B_k$ for temporary storage and keep both these areas
distinct.

\begin{remark}
If we have set associativity ($\ge 2$) in all levels of cache then we
do not need an intermediate buffer $x$ as line $i$ and $j$ can both
reside in cache simultaneously and movement from one to the other will
not cause thrashing. Thus the constant will come down to two.	Since
at any point in time we will only be dealing with two cache lines and
will not need the lines $i$ or $j$ once we have read or written to
them the replacement policy of the cache does not affect our
algorithm.
\end{remark}

\begin{remark}
If the capacity of the register file is greater than the size of the
cache line ($B_k$) of the outermost cache level ($\mathcal{L}_k$) then
we can move data without worrying about collision by copying from line
$i$ to registers and then from registers to line $j$. Thus even in
this case the constant will come down to two.
\end{remark}


Once we have the submatrices in contiguous locations we perform the
transpose as follows.
For each of the submatrices we divide the $B_r \times B_r$ submatrix
(say $S$) in level $\mathcal{L}_r$ (for $2 \le r \le k$) further into
$B_{r-1} \times B_{r-1}$ size submatrices as before. Each $B_{r-1}
\times B_{r-1}$ size subsubmatrix fits into $\mathcal{L}_{r-1}$ cache
completely (since $B_{r-1} \cdot B_{r-1} \leq B_{r-1} \cdot B_k \leq
B_{r-1} \cdot L_{r-1}$ from equation (\ref{bandsize})).
Let $B_r/B_{r-1} = k_r$.

Thus we have the submatrices as 
\[
\left(
\begin{array}{ccccc}
S_{1,1}	       	& S_{1,2}       & \ldots  	& S_{1,k_r}    \\
\vdots         	& \vdots	& \vdots 	& \vdots \\
S_{k_r,1}  	& \ldots	& \ldots 	& S_{k_r,k_r}
\end{array} 
\right) 
\]

So we perform matrix transpose of each $S_{i,j}$ in place without
incurring any misses as it resides completely inside the cache. 
Once we have transposed each $S_{i,j}$ we exchange $S_{i,j}$ with
$S_{j,i}$.
We will show that $S_{i,j}$ and $S_{j,i}$ can not conflict in
$\mathcal{L}_{r-1}$-cache for $i \ne j$.

\begin{figure}[tbp]
\begin{center}
\epsfig{file=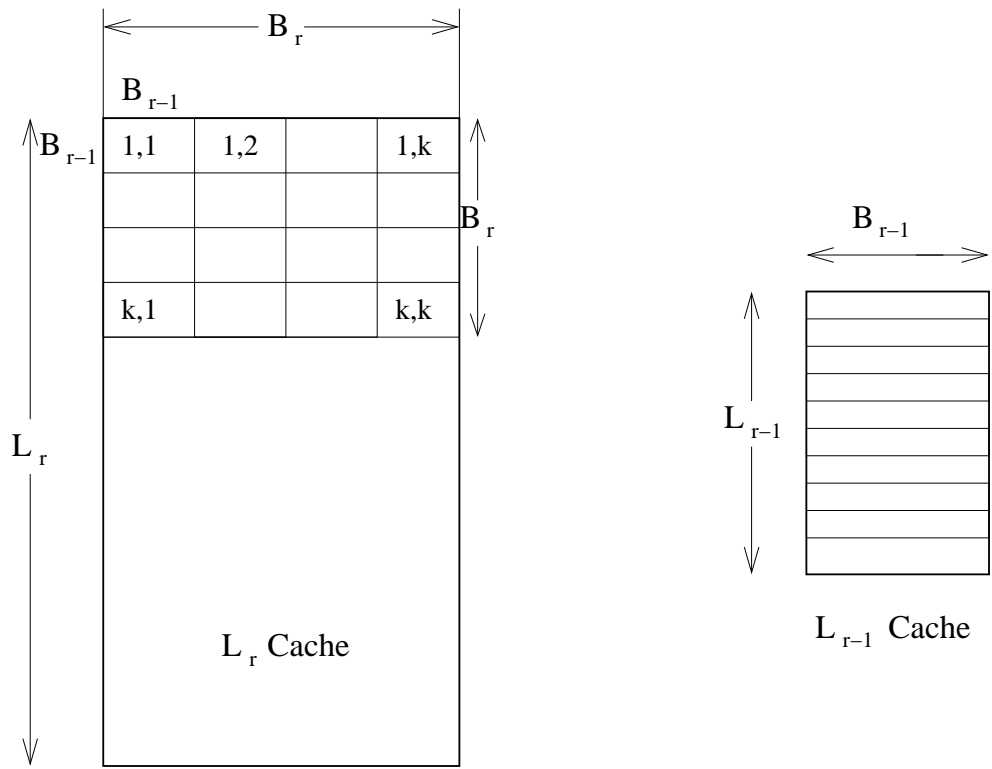,width=\linewidth}
\end{center}
\caption{Positions of symmetric submatrices in Cache}
\label{symmat}
\end{figure}

Since $S_{i,j}$ and $S_{j,i}$ lie in different parts of the
$\mathcal{L}_r$-cache lines, 
they will map to different cache sets in the
$\mathcal{L}_{r-1}$-cache.
The rows of $S_{i,j}$ and $S_{j,i}$ correspond to $({i B_{r-1}} +
a_1)k_r + j$ and $({j B_{r-1}} + a_2)k_r + i$ where $a_1,a_2 \in
\{1,2....B_{r-1}\}$ and 
\[B_r/B_{r-1} = k_r .\] 
If these conflict then
\[ ({i B_{r-1}} + a_1) k_r + j \equiv ({j B_{r-1}} + a_2) k_r + i
(\bmod L_{r-1}).\]
Since $B_{r-1} = 2^u$ and $B_r = 2^v$ and $\mathcal{L}_{r-1} = 2^w$
(all powers of two)
\[ k_r = 2^{v-u} \]
Therefore $k_r$ divides $\mathcal{L}_{r-1}$ (because $k_r =
B_r/B_{r-1} < B_r \le L_{r-1}$).
Hence 
\[j \equiv i (\bmod k_r).\]
Since $i,j \le k_r$ the above implies 
\[i = j.\]

Note that $S_{i,i}$'s do not have to be exchanged.
Thus, 
we have shown that a $B_r \times B_r$ matrix can be divided into
$B_{r-1} \times B_{r-1}$ which completely fits into
$\mathcal{L}_{r-1}$-cache. 
Moreover,
the symmetric sub-matrices do not interfere with each other. 
The same argument can be extended to any $B_j \times B_j$ submatrix
for $j < r$.
Applying this recursively we end up dividing the $B_k \times B_k$ size
matrix in $\mathcal{L}_k$-cache to $B_1 \times B_1$ sized submatrices
in $\mathcal{L}_1$-cache,
which can then be transposed and exchanged easily. 
From the preceding discussion, 
the corresponding submatrices do not interfere in any level of the
cache.

(Note that even though we keep subdividing the matrix at every cache
level recursively and claim that we then have the submatrices in cache
and can take the transpose and exchange them, the actual movement,
i.e., transpose and exchange happens only at the $\mathcal{L}_1$-cache
level, where the submatrices are of size $B_1 \times B_1$.)

The time taken by this operation is 
\[
\sum \frac{ N^2 }{B_i}l_i.
\]

This is because each $S_{i,j}$ and $S_{j,i}$ pair (such that $i \ne
j$) has to be brought into $\mathcal{L}_{r-1}$ Cache only once for
transposing and exchanging of $B_1 \times B_1$ submatrices. 
Similarly, 
at any level of cache, 
a block from the matrix is brought in only once. 
The sequence of the recursive calls ensures that each cache line is
used completely as we move from sub-matrix to sub-matrix.

Finally, 
we move the transposed symmetric submatrices of size $B_k \times B_k$
to their location in memory, \ie,
reverse the process of bringing in blocks of size $B_k$ from random
locations to a contiguous block. 
This procedure is exactly the same as in Theorem \ref{blkmove} in the
previous section that has the constant 3.

\begin{remark}
The above constant of 3 for writing back the matrix to an
appropriate location depends on the assumption that we can keep the
two symmetric submatrices of size $B_k \times B_k$ in contiguous
locations at the same time. 
This would allow us to exchange the matrices during the write back
stage. 
If we are restricted to a contiguous temporary space of size $B_k
\times B_k$ only, 
then we will have to move the data twice, 
incurring the cost twice.
\end{remark}

\begin{remark}
Even though in the above analysis we have always assumed a square
matrix of size $N \times N$ the algorithm works correctly without any
change for transposing a matrix of size $M \times N$ if we are
transposing a matrix $A$ and storing it in $B$. 
This is because the same analysis of subdividing into submatrices of
size $B_k \times B_k$ and transposing still holds. 	
However if we want to transpose a $M \times N$ matrix in place then
the algorithm fails because the location to write back to would not be
obvious and the approach used here would fail.
\end{remark}


\begin{theorem}
	The algorithm for matrix transpose runs in 
\[
O \left(\sum_{i=1}^{i=k} \frac{N^2}{B_i}l_i \right) + O( N^2 )
\]
steps in a computer that has $k$ levels of direct-mapped cache.
\end{theorem}

If we have temporary storage space of size $2 B_k \times B_k + 4 B_k$
and assume block alignment of all submatrices then the constant is 7.
This includes $3$ for initial movement to contiguous location, $ 1$ for 
transposing the symmetric submatrices of size $B_k \times B_k$ and 
$ 3$ for writing back the transposed submatrix to its original location. 
Note that the constant is independent of the number of levels of cache.

\begin{remark}
Even if we have set associativity ($\ge 2$) in any level of cache the
analysis goes through as before (though the constants will come down
for data copying to contiguous locations). For the transposing and
exchange of symmetric submatrices the set associativity will not come
into play because we need a line only once in the cache and are using
only 2 lines at a given time. So either LRU or even FIFO replacement
policy would only evict a line that we have already finished using.
\end{remark}


%
%
%
\subsection{Sorting in multiple levels}
We first consider a restriction of the model described above
where data cannot be transferred simultaneously across non-consecutive
cache levels. We use $C_i$ to denote $\sum_{j=1}^{j=i} M_j$.

\begin{theorem}
The lower bound for sorting in the restricted multi-level cache model
is $\Omega ( N\log N + \sum_{i=1}^k \ell_i \cdot \frac{N}{B_i}
\frac{\log {N/ B_i }}{\log {C_i / B_i }} )$. 
\label{restr}
\end{theorem}

\begin{proof}
The proof of Aggarwal and Vitter can be modified to disregard block
transfers that merely rearrange data in the external memory. 
Then it can be applied separately to each cache level, 
noting that the data transfer in the higher levels do not contribute
for any given level.
\end{proof}
These lower bounds are in the same spirit as those of Vitter and
Nodine~\cite{NV:91} (for the S-UMH model) and Savage~\cite{Sv:95}, 
that is,
the lower bounds do not capture the simultaneous interaction of the
different levels.

If we remove this restriction, 
then the following can be proved along similar lines as Theorem
\ref{lbnd_srt}.
\begin{lemma}
The lower bound for sorting in the multi-level cache model is
\[ \Omega ( \max_{i=1}^k \{ N\log N ,
\ell_i \cdot \frac{N \cdot \log {N/ B _i }}{B_i \log {C_i / B_i }} \}
). \] \noproof
\end{lemma}

This bound appears weak if $k$ is large. 
To rectify this, 
we observe the following. 
Across each cache boundary, 
the minimum number of I/Os follow from Aggarwal and Vitter's
arguments.
The difficulty arises in the multi-level model as a block transfer in
level $i$ propagates in all levels $j < i$ although the block sizes
are different. 
The minimum number of I/Os from (the highest) level $k$ remains
unaffected, 
namely,
$\frac{N}{B_k} \frac{\log {N/ B_k }}{\log {C_k / B_k }}$.  
For level $k-1$, 
we will subtract this number from the lower bound of
$\frac{N}{B_{k-1}} \frac{\log {N/ B_{k-1} }}{\log {C_{k-1} / B_{k-1}
}}$.
Continuing in this fashion, 
we obtain the following lower bound.
\begin{theorem}
\label{glbnd_srt}
The lower bound for sorting in the multi-level cache model is
\[ \Omega \left( N\log N + \sum_{i=1}^{k} \ell_i \cdot \left( 
\frac{N \cdot \log {N/ B_i }}{B_i \log {C_i / B_i }} - \left(
\sum_{j=i+1}^{k} \frac{N \cdot \log {N/ B_j }}{B_j \log {C_j / B_j }}
\right) \right) \right). \]
\noproof
\end{theorem}

If we further assume that $\frac{ C_i }{C_{i-1}} \geq
\frac{B_i}{B_{i-1}} \geq 3$, 
we obtain a relatively simple expression that resembles Theorem
\ref{restr}. 
Note that the consecutive terms in the expression in the second
summation of the previous lemma decrease by a factor of 3. 

\begin{corollary}
The lower bound for sorting in the multi-level cache model with
geometrically decreasing cache sizes and cache lines is
$\Omega ( N\log N + \frac{1}{2}\sum_{i=1}^k  \ell_i \cdot \frac{N
\cdot \log {N/ B_i }}{B_i \log {C_i / B_i }} )$.
\label{nearly}
\noproof
\end{corollary}


\begin{theorem}
In a multi-level cache, 
where the $B_i$ blocks are composed of $B_{i-1}$ blocks,
we can sort in \textbf{expected} time 
$O\left( N\log N  + \left(\frac{\log N/ B_1}{\log M_1 / B_1 } \right)
\cdot\sum_{i=1}^{k} \ell_i \cdot \frac{N}{B_i} \right)$.
\end{theorem}

\begin{proof}
We perform a $M_1 / B_1$-way mergesort using the variation proposed by
Barve \etal~\cite{BGV:96} in the context of parallel disk I/Os. 
The main idea is to shift each sorted stream cyclically by a random
amount $R_i$ for the $i$th stream. 
If $R_i \in [0, M_k -1 ]$,
then the leading element is in any of the cache sets with equal
likelihood.
Like Barve \etal~\cite{BGV:96}, 
we divide the merging into phases where a phase outputs $m$ elements,
where $m$ is the merge degree. 
In the previous section we counted the number of
conflict misses for the input streams, 
since we could exploit symmetry based on the random input. 
It is difficult to extend the previous arguments to a worst case input. 
However, it can be shown easily that if $\frac{m}{s} < \frac{1}{m^3}$ 
(where $s$ is the number of cache sets),
the expected number of conflict misses is $O(1)$ in each phase. 
So the total expected number of cache misses is $O(N/ B_i )$ in the level
$i$ cache for all $1 \leq i \leq k$.

The cost of writing a block of size $B_1$ from level $k$ is spread
across several levels. 
The cost of transferring $B_k /B_1$ blocks of size $B_1$ from level
$k$ is $\ell_k + \ell_{k-1} \frac{B_k}{B_{k-1}} + \ell_{k-2} \frac{
B_{k}}{B_{k-1}} \frac{ B_{k-1}}{B_{k-2}} + \cdots + \ell_1
\frac{B_k}{B_1}$. 
Amortizing this cost over $B_k / B_1$ transfers gives us the required
result. Recall that $O\left( N/ B_1 (\frac{\log N/ B_1}{ \log M_1 / B_1 })
\right)$
$B_1$ block transfers suffice for ${( M_1 / B_1 )}^{1/3}$-way mergesort.
\end{proof}

\begin{remark}
This bound is reasonably close to that of Corollary \ref{nearly} if
we ignore constant factors.  
Extending this to the more general emulation scheme of Theorem
\ref{u_bnd} is not immediate as we require the block transfers across
various cache boundaries to have a nice pattern, 
namely the \emph{sub-block} property.
This is satisfied by the mergesort 
and quicksort 
and a number of other algorithms but cannot be assumed in general.
\end{remark}

\subsection{Cache-oblivious sorting}

In this section, we will focus on a two-level cache model that has
limited associativity.
One of the {\em cache-oblivious} algorithms presented by Frigo
\etal~\cite{FLPR:99} is the funnel sort algorithm. 
They showed that the algorithm is optimal in the I/O model (which is
fully associative). 
However it is not clear whether the optimality holds in the cache
model.  
In this section,
we show that,
with some simple modification, 
funnel sort is optimal even in the direct-mapped cache model.

The funnel sort algorithm can be described as follows.

\begin{itemize}
\item Split the input into ${n}^{1/3}$ contiguous arrays of size
${n}^{2/3}$ and sort these arrays recursively.
\item Merge the ${n}^{1/3}$ sorted sequences using a
${n}^{1/3}$-merger, where a $k$-merger works as follows.
\end{itemize}

\begin{figure}[htbp]
\begin{center}
\epsfig{file=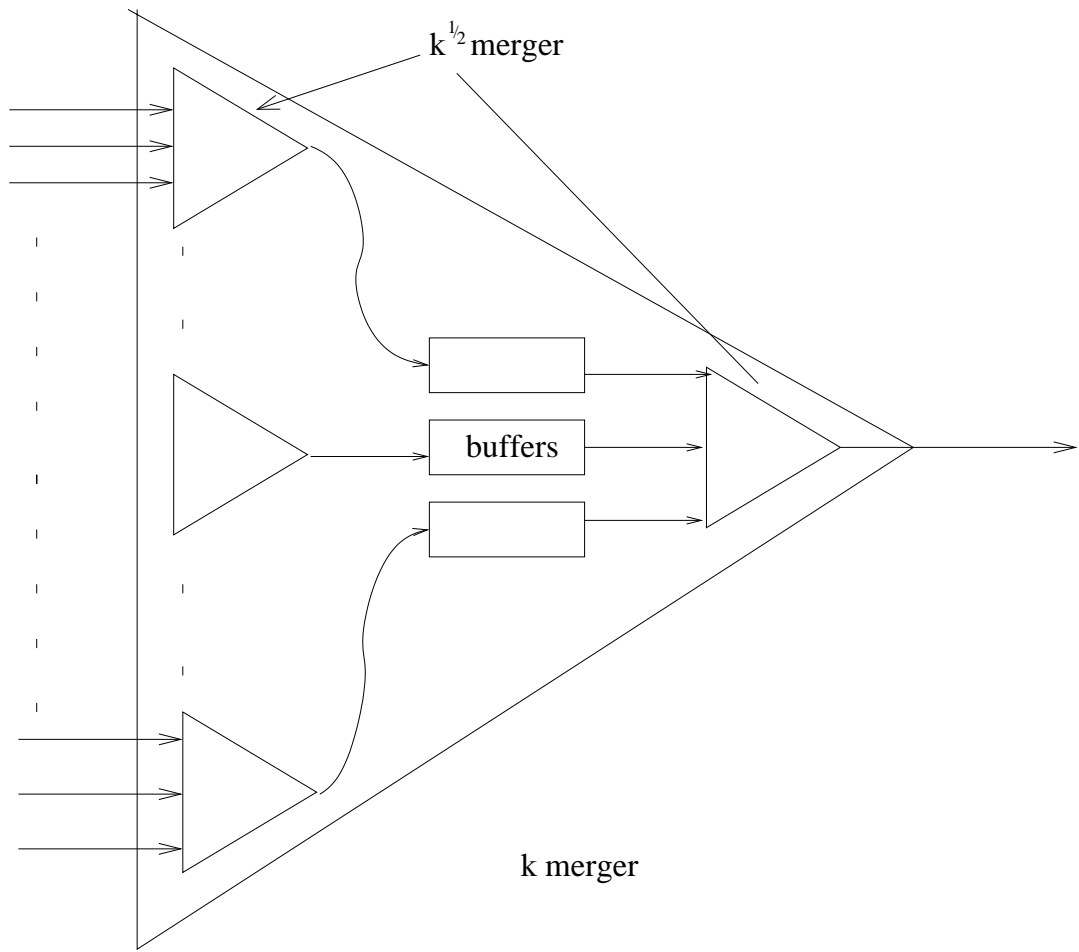,width=\linewidth}
\end{center}
\caption{Recursive definition of a k-merger in terms of $k^{1/2}$-mergers}
\label{kmerger}
\end{figure}

A $k$-merger operates by recursively merging sorted sequences. 
Unlike mergesort,
a $k$-merger stops working on a merging sub-problem when the merged
output sequence becomes ``long enough'' and resumes working on another
merging sub-problem (see Figure \ref{kmerger}).

{\sc Invariant} The invocation of a $k$-merger outputs the first
${k}^{3}$ elements of the sorted sequence obtained by merging the $k$
input sequences.

{\sc Base Case} ${k} = 2$ producing ${k}^{3} = 8$ elements whenever
invoked.

{\sc Note} The intermediate buffers are twice the size of the output
obtained by a ${k}^{1/2}$ merger.

To output ${k}^{3}$ elements, the $k$-merger is invoked ${k}^{3/2}$
times.
Before each invocation the $k$-merger fills each buffer that is less
than half full so that every buffer has at least ${k}^{3/2}$
elements---the number of elements to be merged in that invocation.

Frigo\etal~\cite{FLPR:99} have shown that the above algorithm 
(that does not make explicit use of the various memory-size parameters)
is optimal in the I/O model. 
However,
the I/O model does not account for conflict misses since it assumes
full associativity. 
This could be a degrading influence in the presence of limited
associativity
(in particular direct-mapping).

\subsubsection{Structure of $k$-merger}
It is sufficient to get a bound on cache misses in the cache model
since the bounds for capacity misses in the cache model are the same
as the bounds shown in the I/O model.

\begin{figure}[htbp]
\begin{center}
\epsfig{file=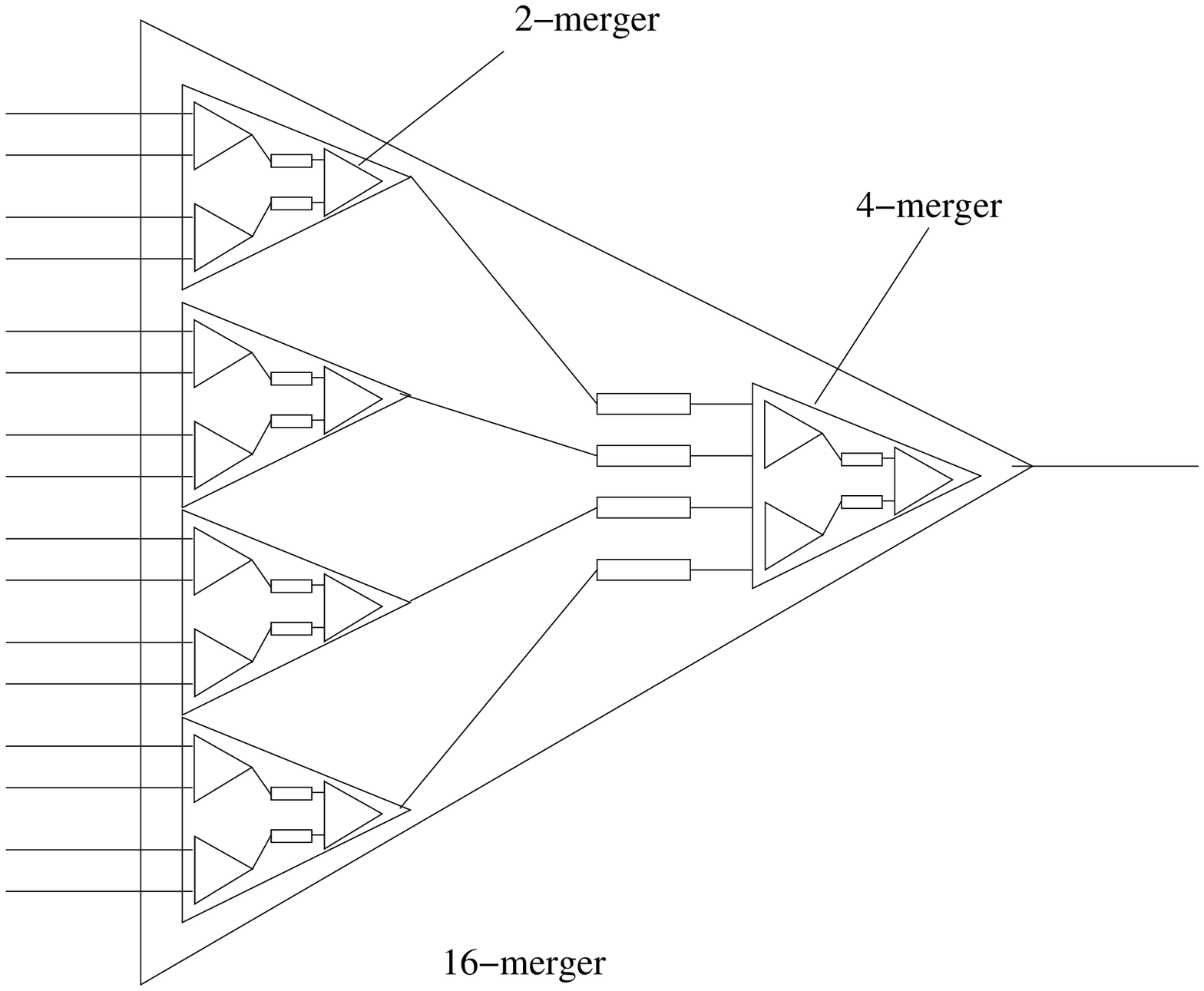,width=\linewidth}
\end{center}
\caption{Expansion of a 16-merger into 2-mergers}
\label{16merger}
\end{figure}

Let us get an idea of what the structure of a $k$-merger looks like by
looking at a 16-merger (see Figure \ref{16merger}).
A $k$-merger, unrolled,
consists of 2-mergers arranged in a tree-like fashion.
Since the number of 2-mergers gets halved at each level and the
initial input sequences are $k$ in number there are $\lg k$ levels.

\begin{lemma}
If the buffers are randomly placed and the starting position is also
randomly chosen (since the buffers are cyclic this is easy to do) the
probability of conflict misses is maximized if the buffers are less
than one cache line long.
\label{confprob}
\end{lemma}

The worst case for conflict misses occurs when the buffers are less
than one cache line in size. This is because if the buffers collide
then all data that goes through them will thrash. If however the size
of the buffers were greater than one cache line then even if some two
elements collide the probability of future collisions would depend
upon the data input or the relative movement of data in the two
buffers.  The probability of conflict miss is maximized when the
buffers are less than one cache line.  Then probability of conflict is
$1/m$, where $m$ is equal to the cache size $M$ divided by the cache
line size $B$, \ie, the number of cache lines.

\subsubsection{Bounding conflict misses}

The analysis for compulsory and capacity misses goes through without change
from the I/O model to the cache model. Thus,
funnel sort is optimal in the cache model
if the conflict misses can be bounded by 
\[
\frac{N}{B} \times \frac{\log {N/B}}{\log {M/B}}
\]

\begin{lemma}
If the cache is 3-way or more set associative, there will be no
conflict misses for a 2-way merger.
\label{setassoc}
\end{lemma}

\begin{proof}
The two input buffers and the output buffer, even if they map to the
same cache set can reside simultaneously in the cache. Since at any
stage only one 2-merger is active there will be no conflict misses at
all and the cache misses will only be in the form of capacity or
compulsory misses.
\end{proof}

\subsubsection{Direct-Mapped case}
	
For an input of size $N$, a ${N}^{1/3}$-merger is created. 
The number of levels in such a merger is $\log {N}^{1/3}$ (
i.e., the number of levels of the tree in the unrolled merger).
\begin{figure}[htbp]
\begin{center}
\epsfig{file=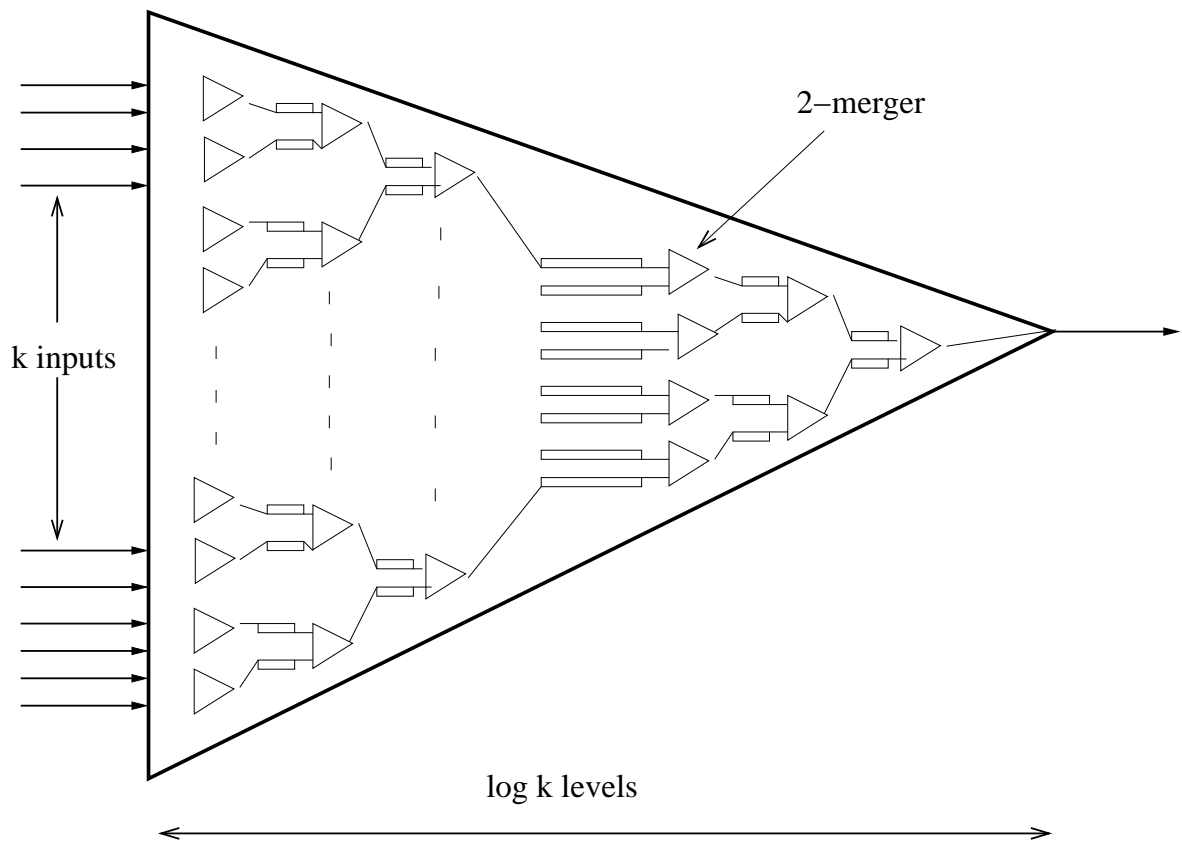,width=\linewidth}
\end{center}
\caption{A k-merger expanded out into 2-mergers}
\label{expkmerger}
\end{figure}
Every element that travels through the ${N}^{1/3}$-merger 
sees $\log {N}^{1/3}$ 2-mergers (see Figure \ref{expkmerger}). For an 
element passing through a 2-merger there are 3 buffers that could collide. 
We {\em charge} an element for a conflict miss if it is swapped out of
the cache before it passes to the output buffer or collides
with the output buffer when it is being output. 
So the expected number of collisions is $^3C_2$ times the probability of 
collision between any two buffers (two input and one output). 
Thus the expected number of collisions 
for a single element passing through a 2-merger is $^3C_2 \times 1/m \le 3/m$ 
where $m = M/B$. 

If $x_{i,j}$ is the probability of a cache miss for element $i$ in level $j$ then summing over all elements and all levels we get 
\begin{eqnarray*}
E \left( \sum_{i=1}^{N} \sum_{j=1}^{N^{1/3}} x_{i,j} \right) & = & \sum_{i=1}^{N} \sum_{j=1}^{\log {N^{1/3}}} E (x_{i,j}) \\
& \le & \sum_{i=1}^{N} \sum_{j=1}^{\log {N^{1/3}}} \frac{3}{m} = \frac{3N}{m} \times \log {N^{1/3}} \\
& = & O\left( \frac{N}{m} \times \log {N}\right)
\end{eqnarray*}

\begin{lemma}
The expected performance of funnel sort is optimal in the 
direct-mapped cache model if
$\log \frac{M}{B} \le \frac{M}{B^2\log B}$. 
It is also optimal for a 3-way associative cache.
\end{lemma}

\begin{proof}
If $M$ and $B$ are such that
\[
\log \frac{M}{B} \le \frac{M}{B^2 \log B}
\]
we have the total number of conflict misses
\[
\frac{N \log N}{m} = \frac{N \log N}{B \log B \frac{M}{B^2 \log B}} 
\le \frac{N}{B} \times 
\frac{\log {N/B}}{\log {M/B}}
\]
Note that the condition is satisfied for $M \geq B^{2 + \epsilon}$ for any fixed
$\epsilon > 0$ which is similar to the {\em tall-cache} assumption
made by Frigo \etal.

The set associative case is proved by Lemma \ref{setassoc}.
\end{proof}

The same analysis is applicable between successive levels $\mathcal{L}_i$
and $\mathcal{L}_{i+1}$ of a multi-level cache model. 
This yields an optimal algorithm for sorting
in the multilevel cache model.
\begin{theorem}
In a multi-level cache model, the number of cache misses at level 
$\mathcal{L}_i$ in the funnel sort algorithm 
can be bounded by $\frac{N \log(N/ B_i )}{B_i \log (M_i /B_i )}$. 
\end{theorem}
This bound matches the lower bound of Lemma \ref{restr} within a constant
factor,
which makes it an optimal algorithm when simultaneous transfers are
not allowed across multiple levels.


\section{Conclusions}
\LABEL{sec}{concl}

We have presented a cache model for designing and analyzing
algorithms.
Our model,
while closely related to the I/O model of Aggarwal and Vitter,
incorporates three additional salient features of cache:
lower miss penalty,
limited associativity,
and lack of direct program control over data movement.
We have established an emulation scheme that allows us to
systematically convert an I/O-efficient algorithm into a
cache-efficient algorithm.
This emulation provides a generic starting point for cache-conscious
algorithm design;
it may be possible to further improve cache performance by
problem-specific techniques to control interference misses.
We have also demonstrated the relevance of the emulation scheme by
demonstrating that a direct mapping of an I/O-efficient algorithm does
not guarantee a cache-efficient algorithm.
Finally, we have extended our basic cache model to multiple cache levels.

Our single-level cache model is based on a blocking direct-mapped
cache that does not distinguish between reads and writes.
Modeling a non-blocking cache or distinguishing between reads and
writes would appear to require queuing-theoretic extensions and does
not appear to be appropriate at the algorithm design stage.
The \emph{translation lookaside buffer} or TLB is another important
cache in real systems that caches virtual-to-physical address
translations.
Its peculiar aspect ratio and high miss penalty raise different
concerns for algorithm design.
Our preliminary experiments with certain permutation problems suggests
that TLBs are important to model and can contribute significantly to
program running times.

We have begun to implement some of these algorithms to validate the
theory on real machines, and also
using cache simulation tools like
\emph{fast-cache}, \textsc{atom}, or \emph{cprof}. Preliminary
observations indicate that our predictions are more accurate with
respect to miss ratios than actual running times
(see~\cite{chatterjee00}).
We have traced a number of possible reasons for this. First,
because the cache miss latencies are not astronomical,
it is important to keep track of the constant factors.
An algorithmic variation that guarantees lack of conflict misses
at the expense of doubling the number of memory references may turn
out to be slower than the original algorithm.
Second,
our preliminary experiments with certain permutation problems suggests
that TLBs are important to model and can contribute significantly to
program running times.
Third,
several low-level details hidden by the compiler related to
instruction scheduling,
array address computations,
and alignment of data structures in memory can significantly influence
running times. 
As argued earlier, these factors are more appropriate to tackle at the
level of implementation than algorithm design.

Several of the cache problems we observe can be traced to the simple
array layout schemes used in current programming languages.
It has shown
elsewhere~\cite{chatterjee99a,chatterjee99b,thottethodi98} that
nonlinear array layout schemes based on quadrant-based decomposition
are better suited for hierarchical memory systems.
Further study of such array layouts is a promising direction for
future research.

%


\section*{Acknowledgments}

We are grateful to Alvin Lebeck for valuable discussions related to
present and future trends of different aspects of memory hierarchy
design.  
We would like to acknowledge Rakesh Barve for discussions related to
sorting, FFTW, and BRP. 
The first author would also like to thank Jeff Vitter for his comments
on an earlier draft of this paper.

\appendix
\section{Approximating probability of conflict}
\LABEL{app}{approx}

Let $\mu$ be the number of elements between $S_{i,j}$ and $ S_{i,j+1}$,
\ie, 
one less than the difference in ranks of $S_{i,j}$ and $ S_{i,j+1}$.
($\mu$ may be 0, 
which guarantees event E1.)
Let $E_m$ denote the event that $\mu = m$. 
Then $\Pr [ E1 ] = \sum_m \Pr [ E1 \cap E_m ] $, 
since $E_m$'s are disjoint. 
For each $m$,
$ \Pr [ E1 \cap E_m ] = \Pr [ E1 | E_m ] \cdot \Pr [ E_m ]$. 
The events $E_m$ correspond to a geometric distribution, 
\ie,
\begin{equation}
\Pr[E_m] = \Pr [ \mu = m] = \frac{1}{k} {\left( 1 - \frac{1}{k}
\right)}^{m}.
\label{geomdist}
\LABEL{eqn}{geomdist}
\end{equation}

To compute $ \Pr [ E1 | E_m ] $, 
we further subdivide the event into cases about how the $m$
numbers are distributed into the sets $S_j , j \neq i$.
Wlog, 
let $i=1$ to keep notations simple. 
Let $m_2, \ldots, m_k$ denote the case that $m_j$ numbers belong to
sequence $S_j$  ($\sum_j m_j = m$). 
We need to estimate the probability that for sequence $S_j$, 
$b_j$ does not conflict with $\BS ( b_1)$ 
(recall that we have fixed $i=1$) 
during the course that $m_j$ elements arrive in $S_j$. 
This can happen only if $\BS ( b_j )$ 
(the cache set position of the leading block of $S_j$ right after
element $S_{1,t}$) 
does not lie roughly $\lceil m_i / B \rceil$ blocks from $\BS ( b_1 )$.
From assumption A1 and some careful counting this is $1 - \frac{ m_j
-1 +B}{s B}$ for $m_j \geq 1$.
For $m_j = 0$, 
this probability is 1 since no elements go into $S_j$ and hence
there is no conflict.\footnote{The reader will soon realize
that this case leads to some non-trivial calculations.}
These events are independent from our assumption A1 and hence these
can be multiplied. 
The probability for a fixed partition $m_2 , \ldots, m_k$ is the
multinomial
$\frac{m !}{ m_2 ! \cdots m_k !} \cdot 
{\left(\frac{1}{k-1} \right)}^m$ 
($m$ is partitioned into $k-1$ parts).
Therefore we can write the following expression for $\Pr [ E1 | E_m ]$.
\begin{equation}
\Pr [ E1 | E_m ] = \sum_{ m_2 + \cdots + m_k = m} 
\frac{m !}{ m_2 ! \cdots m_k !} \cdot
{\left(\frac{1}{k-1} \right)}^m \prod_{m_i \neq 0} \left( 1 - 
\frac{ m_j -1 +B}{s B} \right)
\label{eqncond}
\LABEL{eqn}{eqncond}
\end{equation}

In the remainder of this section,
we will obtain an upper bound on the right hand side of \eqn{eqncond}.
Let $nz( m_2, \ldots, m_k )$ denote the number of $j$s for which
$m_j \neq 0$ (non-zero partitions). 
Then \eqn{eqncond} can be rewritten as the following
inequality.
\begin{equation}
\Pr [ E1 | E_m ] \leq \sum_{ m_2 + \cdots + m_k = m}
\frac{m !}{ m_2 ! \cdots m_k !} \cdot
{\left(\frac{1}{k-1} \right)}^m {\left( 1 -
\frac{ 1}{s } \right)}^{nz( m_2 \ldots m_k )}
\label{ineqcond}
\LABEL{eqn}{ineqcond}
\end{equation}
since $ \left( 1 - \frac{ m_j -1 +B}{s B} \right) \leq \left( 1 -
\frac{ 1}{s } \right)$ for $m_j \geq 1$.
In other words,
the right side is the expected value of  
${\left( 1 - \frac{ 1}{s } \right)}^{NZ( m,k-1 )}$,
where $NZ( m,k-1 )$ denotes the number of non-empty bins when $m$
balls are thrown into $k-1$ bins.
Using \eqn{geomdist} and the preceding discussion,
we can write 
down an upper bound for the (unconditional) probability of $E1$ as
\begin{equation}
\sum_{m=0}^{\infty}  \frac{1}{k} {\left( 1 - \frac{1}{k}
\right)}^{m} \cdot E \left[ {\left( 1 - \frac{ 1}{s } 
\right)}^{NZ( m, k-1 )} \right]
\label{simpeqn}
\LABEL{eqn}{simpeqn}
\end{equation}

We use known sharp concentration bounds for the occupancy problem
to obtain the following approximation for the expression
\eqnref{simpeqn} in terms of $s$ and $k$.
\begin{theorem}[\cite{KMPS:94}]
Let $r = m/n$, 
and $Y$ be the number of empty bins when $m$ balls are thrown randomly
into $n$ bins. 
Then 
\[ E [ Y ] = n {\left( 1 - \frac{1}{m} \right)}^{m} \sim n e^{-r} \]
and for $\lambda > 0$
\[ \Pr [ | Y - E[Y] | \geq \lambda ] \leq
 2 \exp \left( -\frac{\lambda^2 (n-1)/2 }{n^2 - \mu^2 }\right). \]
\label{nzbins}
\LABEL{eqn}{nzbins}
\noproof
\end{theorem}

\begin{corollary}
Let $NZ$ be the number of non-empty bins when $m$ balls are thrown
into $k$ bins. 
Then
\[ E[  NZ ] = k ( 1- e^{-m/k} ) \]
and
\[ \Pr [ | NZ - E[NZ] | \geq \alpha\sqrt{2 k \log k} ] \leq 1/k^\alpha.
\]
\noproof
\end{corollary}
So in \eqn{ineqcond}, 
$E [ {\left( 1 - \frac{ 1}{s } \right)}^{NZ( m, k-1 )}]$ can be
bounded by 
\begin{equation}
 1/k^{\alpha} \left( 1 - 1/s \right) +
{\left( 1 - \frac{ 1}{s } \right)}^{ k ( 1- e^{-m/k} -
     \alpha\sqrt{2 k \log k} /k)}
\label{rghbnd}
\LABEL{eqn}{rghbnd}
\end{equation}
for any $\alpha$ and $m \geq 1$.

\begin{proof}
(of Lemma \ref{weakbnd}): 
We will split up the summation of \eqnref{simpeqn} into two parts,
namely,
$m \leq e/2\cdot k$ and $m > e/2 \cdot k$.
One can obtain better approximations by refining the partitions,
but our objective here is to demonstrate the existence of $\epsilon$
and $\delta$ and not necessarily obtain the best values.

\begin{eqnarray}
\sum_{m=0}^{\infty}  \frac{1}{k} {\left( 1 - \frac{1}{k} \right)}^{m} \cdot
E [ {\left( 1 - \frac{ 1}{s } \right)}^{NZ( m, k-1 )}] &
= & \sum_{m=0}^{ek/2k}  \frac{1}{k} {\left( 1 - \frac{1}{k} \right)}^{m} \cdot
E [ {\left( 1 - \frac{ 1}{s } \right)}^{NZ( m, k-1 )}]\nonumber \\
&+ &  \sum_{m=ek/2+1}^{\infty}  \frac{1}{k} 
{\left( 1 - \frac{1}{k} \right)}^{m} \cdot
E [ {\left( 1 - \frac{ 1}{s } \right)}^{NZ( m, k-1 )}]
\label{aptwoparts}
\LABEL{eqn}{aptwoparts}
\end{eqnarray}

The first term can be upper bounded by
\[ \sum_{m=0}^{ek/2}  \frac{1}{k} {\left( 1 - \frac{1}{k} \right)}^{m} \]
which is $\sim 1 - \frac{1}{ e^{e/2 }} \sim 0.74$.

The second term can be bounded using \eqn{rghbnd} using $\alpha \geq
2$.
\begin{eqnarray}
 \sum_{m=ek/2+1}^{\infty}  \frac{1}{k} 
{\left( 1 - \frac{1}{k} \right)}^{m} \cdot
E [ {\left( 1 - \frac{ 1}{s } \right)}^{NZ( m, k-1 )}]
& \leq &\sum_{m=ek/2+1}^{\infty}  \frac{1}{k} {\left( 1 - \frac{1}{k} \right)}^{m}
\cdot 1/k^2 \left( 1 - 1/s \right) \nonumber\\
& + &
\sum_{m=ek/2+1}^{\infty}  \frac{1}{k} {\left( 1 - \frac{1}{k} \right)}^{m}
\cdot {\left( 1 - \frac{ 1}{s } \right)}^{ k ( 1- e^{-m/k} -
     \alpha\sqrt{2 k \log k} /k)}
\label{apeqn2}
\LABEL{eqn}{apeqn2}
\end{eqnarray}

The first term of the previous equation is less than $1/k$ and the
second term can be bounded by
\[ \sum_{m=ek/2+1}^{\infty}  \frac{1}{k} {\left( 1 - \frac{1}{k} \right)}^{m}
\cdot {\left( 1 - \frac{ 1}{s } \right)}^{0.25 k} \]
for sufficiently large $k$ ($k > 80$ suffices).
This can be bounded by $\sim 0.25 e^{-0.25k/s}$, 
so \eqn{apeqn2} can be bounded by $1/k +  0.25 e^{0.25k/s}$. 
Adding this to the first term of \eqn{aptwoparts}, 
we obtain an upper bound of $ 0.75 +  0.25 e^{-0.25k/s}$ for $k >
100$. 
Subtracting this from 1 gives us $ \frac{1 -  e^{-0.25k/s} }{4}$, 
\ie, $\delta \geq \frac{1 -  e^{-0.25k/s} }{4}$. 
\end{proof}

\end{document}